\documentclass[aip, reprint]{revtex4-2}
\usepackage[utf8]{inputenc}
\usepackage[T1]{fontenc}
\usepackage{epsfig}
\usepackage{float}
\usepackage{mathptmx}
\usepackage{dcolumn}
\usepackage{bm}
\usepackage{graphicx}

\usepackage{times}
\usepackage{float}
\usepackage{amssymb}
\usepackage{amsmath}
\usepackage{amsfonts}
\usepackage{gensymb}
\usepackage{enumitem}
 
\usepackage{setspace}
\usepackage[dvipsnames]{xcolor}
\usepackage{upgreek}
\usepackage{mathtools}
\usepackage[sort&compress]{natbib}

\setcitestyle{open={},close={},numbers,comma,super}

\usepackage[pagebackref=false]{hyperref}
\hypersetup{
    colorlinks  =   true,
    linkcolor   =   BlueViolet,
    citecolor   =   BlueViolet,
    filecolor   =   BlueViolet,
    urlcolor    =   BlueViolet}
\usepackage[capitalise,nameinlink]
{cleveref}
\usepackage{hypcap} 
\usepackage{titlesec}

\titleformat{\section}[hang]{\small\bfseries\sffamily}{\thesection.}{0.5em}{\MakeUppercase}
\titlespacing{\section}{0pc}{1.2pc}{0.3pc}
\titlespacing{\subsection}{0pc}{1pc}{0.2pc}

\makeatletter
\renewcommand*{\fnum@figure}{{\normalfont\bfseries \figurename~\thefigure}}
\renewcommand*{\@caption@fignum@sep}{\textbf{ : }}
\makeatother

\begin{document}

\title{Domain control and periodic poling of epitaxial ScAlN}

\newcommand{\CMT}[1]{{}}

\author{Fengyan Yang}
\affiliation{Department of Electrical Engineering, Yale University, New Haven, CT 06511, USA}
\author{Guangcanlan Yang}
\affiliation{Department of Electrical Engineering, Yale University, New Haven, CT 06511, USA}
\author{Ding Wang}
\affiliation{Department of Electrical Engineering and Computer Science, University of Michigan, Ann Arbor, MI 48109, USA}
\author{Ping Wang}
\affiliation{Department of Electrical Engineering and Computer Science, University of Michigan, Ann Arbor, MI 48109, USA}
\author{Juanjuan Lu}
\affiliation{Department of Electrical Engineering, Yale University, New Haven, CT 06511, USA}
\author{Zetian Mi}
\affiliation{Department of Electrical Engineering and Computer Science, University of Michigan, Ann Arbor, MI 48109, USA}
\author{Hong X. Tang}
\email{hong.tang@yale.edu}
\affiliation{Department of Electrical Engineering, Yale University, New Haven, CT 06511, USA}

\begin{abstract}

ScAlN is an emerging ferroelectric material that possesses large band gap, strong piezoelectricity, and holds great promises for enhanced $\chi^{(2)}$ nonliearity. In this study, we demonstrate high-fidelity ferroelectric domain switching and periodic poling of Al-polar ScAlN thin film epitaxially grown on c-axis sapphire substrate using gallium nitride as a buffer layer.  Uniform poling of ScAlN with periods ranging from 2\,$\upmu$m to 0.4\,$\upmu$m is realized. The ability to lithographically control the polarization of epitaxial ScAlN presents a critical advance for its further exploitation in ferroelectric storage and nonlinear optics applications.

\end{abstract}

\maketitle


Alloying scandium (Sc) with aluminum nitride (AlN) results in an innovative class of ferroelectric nitride semiconductor, Sc$_x$Al$_{1-x}$N, that holds great potential for electronic and photonic applications. Single crystalline ScAlN can be grown on c-axis sapphire substrate with gallium nitride (GaN) as a buffer layer via molecular beam epitaxy  \cite{MBE_grow_ScAlN, wang_MBEgrow}. The Sc composition is tunable which can be leveraged for varying bandgap and refractive index. ScAlN possesses unique properties such as high dielectric strength, large piezoelectric constants, and low coercive fields in comparison to AlN \cite{Moram_review,Rassay2021_lamb_mode_resonator}. These properties make ScAlN attractive for various applications in electronics, including heterostructrue field effect transistors (HFETs), piezoelectric and ferroelectric devices, and high-frequency acoustic resonators\cite{Moram_calculation_piezo_polarization,wangping_npolar,Fichtner_ferror, DingW_ScAlNmemory,Giribaldi2020-yr}. 

Especially, ScAlN promises a high second-order nonlinear optical susceptibility $\chi^{(2)}$, which was reported to increase at higher Sc concentration and can be nearly two times as large as that of LiNbO$_3$ when Sc concentration reaches 20$\%$ \cite{Yoshioka2021-fi}. This particular property leads to enhanced $\chi^{(2)}$ nonlinear interaction, for example, in second harmonic generators\cite{alex_shg, juanjuan_poling, chen2019ultra} and optical parametric oscillators\cite{alex_opo, juanjuan_OPO}. Moreover, the Sc alloying in AlN significantly decrease the coercive field to below the dielectric breakdown field, making it possible to flip the polarization\cite{DingW_ScAlNmemory}. Therefore, if periodically flipping and maintaining the polarization is achievable on a ScAlN platform, quasi-phase matching for $\chi^{(2)}$ process can be realized, further enhancing the effective coupling strength and facilitating numerous nanophotonic applications such as $\chi^{(2)}$ frequency conversion\cite{juanjuan_poling,juanjuan_OPO, chen2019ultra}, Pockels combs\cite{bruch2021pockels}, and cascaded $\chi^{(2)}$ and $\chi^{(3)}$ nonlinearities. \cite{cui2022situ, liu2019beyond,comb_SHG,comb_SHG2,wjq_five_wave} 

This article presents a study on the ferroelectric switching property of epitaxial Al-polar ScAlN thin film grown on GaN-buffered sapphire template, which is highly compatible with standard III-nitrides semiconductor fabrication process. The ferroelectric switching current was extracted using a custom setup based on Positive-Up-Negative-Down (PUND) measurement\cite{Martin2017-bo,pund_HoMnO3}, and the coercive field was found to be 6\,MV/cm. We further demonstrate high fidelity and uniform ScAlN poling with periods ranging from 2\,$\upmu$m to 0.4\,$\upmu$m. Reaching such short poling periods is a prerequisite for second harmonic generation in the deep visible even ultra-violet wavelength range, which remains to be challenging due to stringent requirement on poling periods \cite{Ayed_blue_light,LN_submicron_bipolar, Zhang2020-ty}. The submicron poling period demonstrated in this article provides an opportunity for mirrorless OPO \cite{MOPO_1966, backwardOPO, mirrorlessOPO}, a type of optical parametric oscillator that does not require an optical cavity, which simplifies the design and makes the device more robust against environmental disturbance.

\begin{figure*}[h]
\centering
\includegraphics[width=0.75\linewidth]{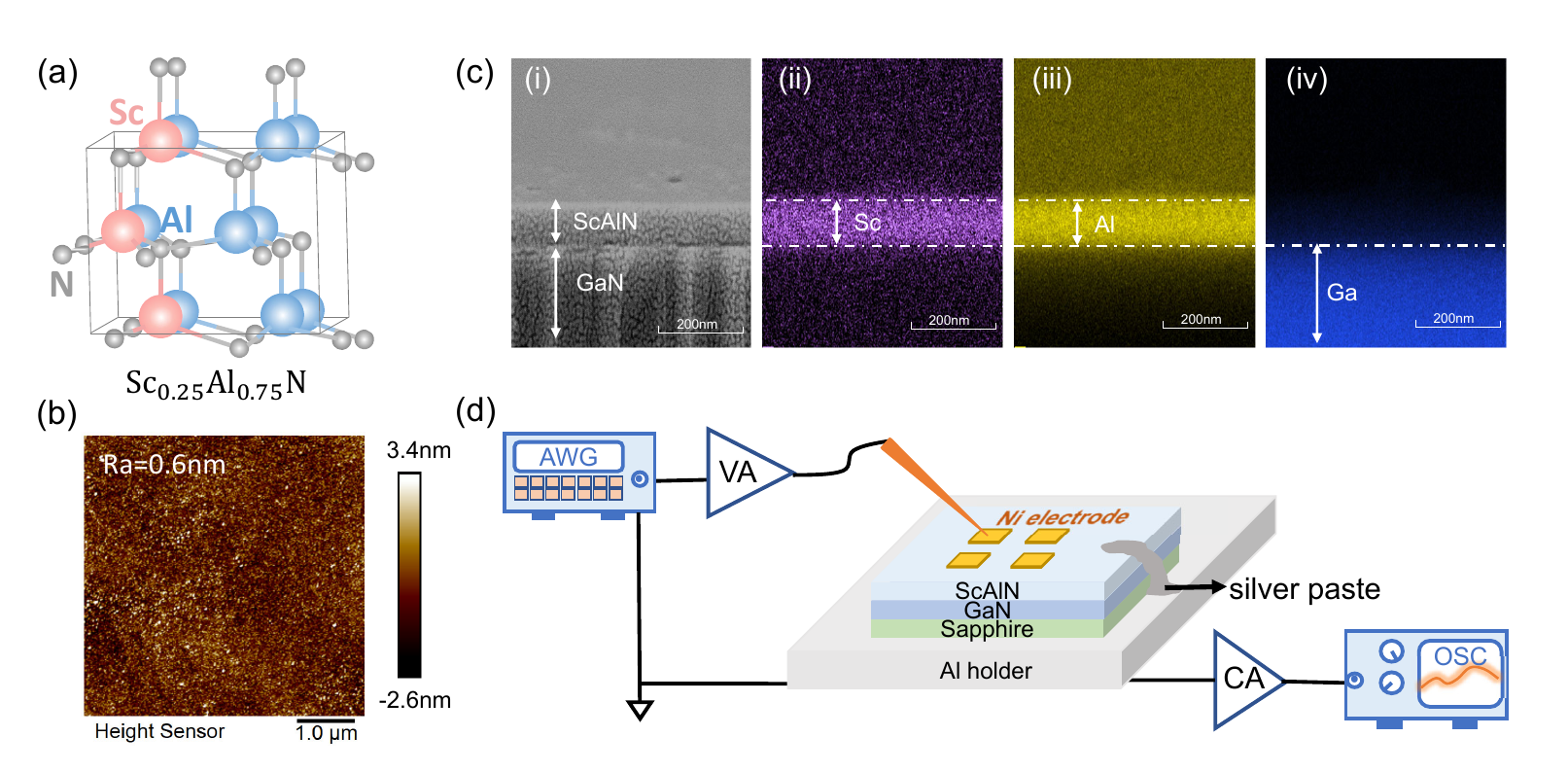}
\caption{ (a) Wurtzite crystal stucture of Al-polar Sc$_{0.25}$Al$_{0.75}$N. (b) AFM image of as-grown Sc$_{0.25}$Al$_{0.75}$N surface across 5$\mathrm{\mu m} \times$ 5$\mathrm{\mu m}$ area with RMS=0.6nm. (c) i, SEM image of Sc$_{0.25}$Al$_{0.75}$N/GaN heterostructure. ii-iv, EDX mapping of Sc, Al, Ga distribution. (d) Experimental setup for PUND measurement and poling. AWG: arbitrary waveform generator, VA: linear voltage amplifier with fixed gain to be 100, CA: current amplifier with variable gain, OSC:oscilloscope.}
\label{figure1}

\end{figure*}

The starting wafer consists of a 100\,nm-thick Al-polar ScAlN film grown on a c-axis sapphire substrate via Molecular Beam Epitaxy (MBE) using highly Si-doped GaN as a buffer layer \cite{MBE_grow_ScAlN}. Fig.~\ref{figure1}(a) shows the wurtzite crystal structure of Al-polar ScAlN alloyed with 25$\%$ Sc concentration, which was set to be our targeted concentration in growth to allow better lattice match to GaN and reduce dislocation density\cite{wangping_npolar}. An atomic-force-microscopy (AFM) scan over a 5\,\textmu m$\times$5\,\textmu m area reveals a smooth surface with a root-mean-squared roughness as low as 0.6\,nm, as indicated in Fig.~\ref{figure1}(b). The layer composition of the wafer was characterized by a scanning electron microscope (SEM) equipped with energy-dispersive X-ray spectroscopy (EDS). Fig.~\ref{figure1}(c) shows the cross-sectional SEM images of of Sc$_{0.25}$Al$_{0.75}$N/GaN heterostructure. The element distribution was obtained by EDS mapping at 4.088\,keV, 1.486\,keV, and 1.098\,keV, which correspond to emission lines of Sc K$\alpha$, Al K$\alpha$, and Ga L$\alpha$, respectively, indicating well defined boundaries at all growth interfaces.

The ferroelectric switching property of ScAlN was characterized by PUND measurement, where a series of programmed voltage pulses was applied to the film successively to extract ferroelectric switching current $I_f$. Electrodes of varying designs were patterned on ScAlN chip through nickel evaporation, followed by a liftoff process. The underlying GaN layer, which has low resistivity due to high Si doping concentration, serves as the counter electrode. Silver paste was applied at the edge of the chip, connecting the Si-doped GaN layer with an aluminum holder. Fig.~\ref{figure1}(d) illustrates the PUND measurement setup. An arbitrary waveform generator (AWG) was programmed to generate PUND pulses, which were amplified by a linear voltage amplifier, and then applied to nickel electrodes via an ultra-sharp probe tip. The current induced by the electric field was converted to voltage signal through a low-noise current amplifier, and an oscilloscope was used to measure the output voltage.   

\begin{figure*}[h]
\centering
\includegraphics[width=0.75\linewidth]{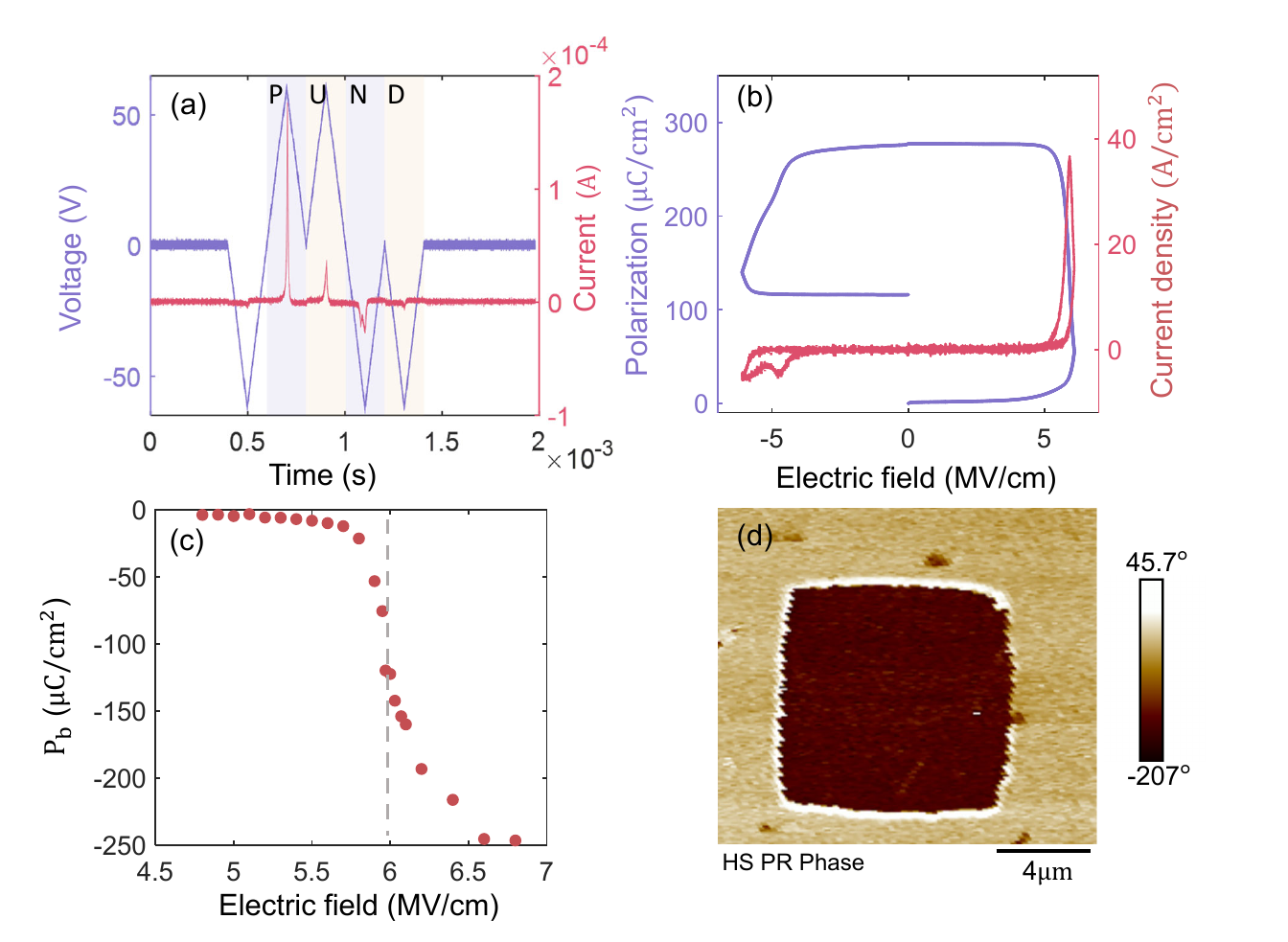}
\caption{(a) Applied voltage pulses and corresponding circuit current in PUND measurement. (b) Forward and backward switching current density (red curve) and P-E loop by integrating $J_f$ with time (purple curve). (c) Backward switched polarization $P_b$ changes with peak electric field. (d) PFM phase image of a 10\textmu m $\times$ 10\textmu m square which was poled by trapezoid voltage pulses with a peak voltage of 62 V and pulse width of 500 \textmu s.}
\label{figure2}
\end{figure*}

Fig.~\ref{figure2}(a) displays the applied voltage pulses and circuit current in PUND measurement. The triangular-shaped voltage pulses have a peak voltage of $\pm$62\,V and a pulse length of 200\,\textmu s.  A negative voltage pulse was applied to initialize the polarization of the film, followed by positive (`P') and up (`U') pulses. In addition to the polarization switching current $I_f$, parasitic currents including displacement current $I_c$ =\,$C\,\frac{dV}{dt}$ and static leakage current $I_l$, contribute to the measured $I_p$. Since polarization was supposed to be fully reversed during "P" pulse which was chosen with a relatively long duration, the ferroelectric current should not present in $I_u$. Therefore, the net forward switching current $I_f$ can be obtained by subtracting $I_u$ from $I_p$. Similarly, the backward switching current can be extracted from negative (`N') and down (`D') pulses, where $I_b=I_n-I_d$.

The ferroelectric current density can be calculated by $J=I/A$, where $A$ is the area of the nickel electrode. The electric field applied on ScAlN film can be approximated as $E=V/d$, with $d=100$\,nm. The time-dependent polarization $P$ of the film can be found by numerically integrating $J$ over time. The resulting $J-E$ and $P-E$ relationships are shown in Fig.~\ref{figure2}(b), represented by red and purple curves, respectively. The $J_{f}$ and $J_{b}$ values exhibit large asymmetry, which is mainly attributed to the asymmetric electrode configurations and the enhanced leakage current $I_v$ during forward switching, as reported in previous work \cite{asymmetry_ref}.

In order to obtain reliable measurements, we focused on the backward switched polarization $P_b$ and swept the peak voltage with fixed pulse length. The results are shown in Fig.\,\ref{figure2}(c). The ScAlN film under investigation exhibited a coercive field of approximately 6\,MV/cm and a maximum backward switched polarization of 250 \textmu C/cm$^2$. Remnant polarization of this film is thus estimated to be $P_{b,max}/2=125~$\textmu C/cm$^2$.

We employed trapezoid voltage pulses with a peak voltage of 62$\,$V and pulse width of 500 \textmu s to perform poling process, which fully reverse the ferroelectric polarization from Al-polar to N-polar. After poling, the top nickel electrodes were removed by hydrochloric acid, then  Piezoresponse Force Microscopy (PFM) was utilized to map out N-polar and Al-polar domains, as shown in Fig.\ref{figure2}\,(d). The significant phase difference between domains confirms the effective poling of ScAlN thin film.

\begin{figure*}[h]
\centering
\includegraphics[width=0.75\linewidth]{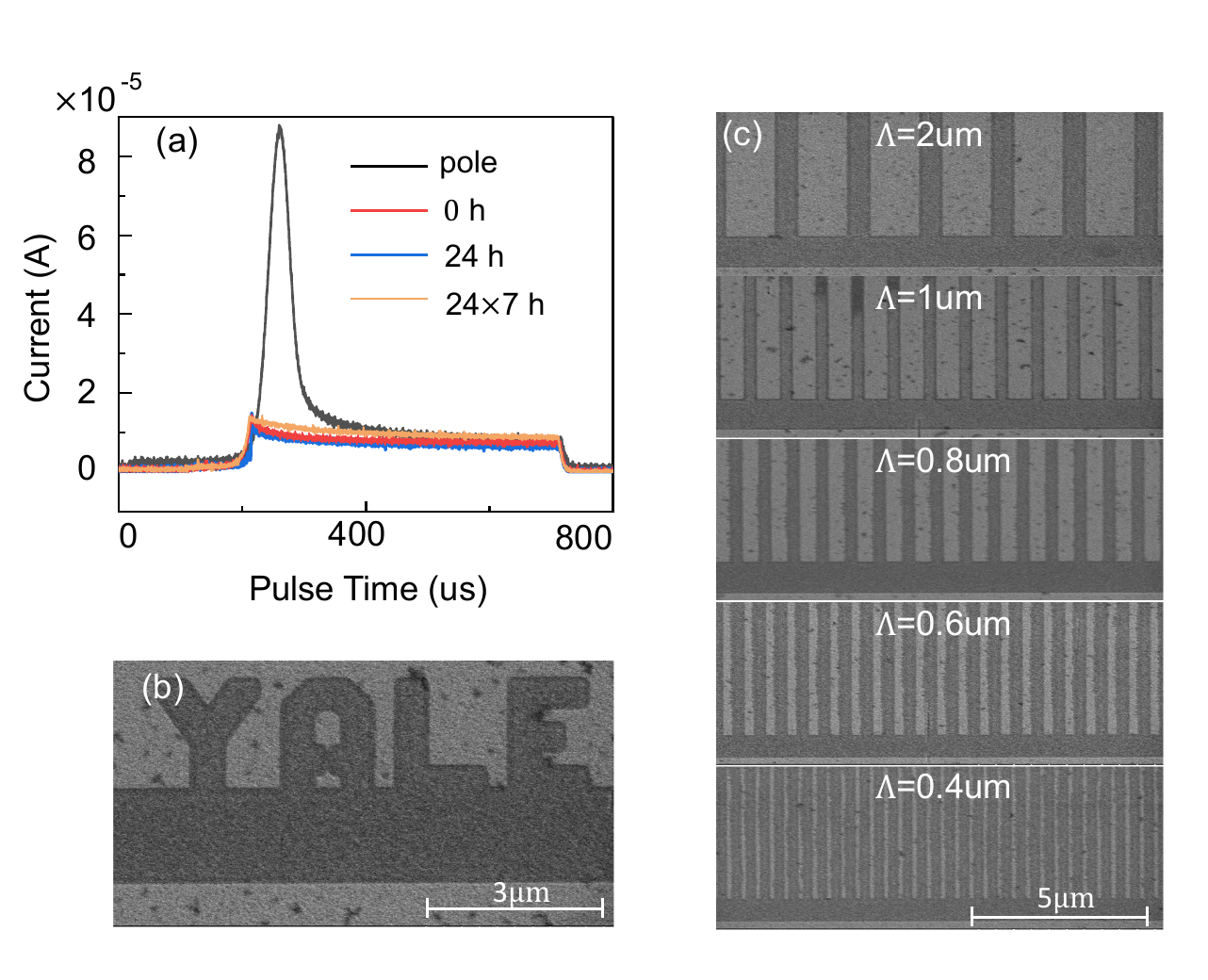}
\caption{(a) Currents for the poling pulse and for the probing pulse at different time. (b) SEM image of a "YALE" pattern after poling and HCl selective etching. (c) SEM images of periodical poling pattern with periods equal to 2, 1, 0.8, 0.6, 0.4\,$\upmu$m, respectively.}
\label{figure3}
\end{figure*}

To verify the polarization retention capability of ScAlN film, we first applied a poling voltage pulse to a selected device. After certain periods of time time, we repeated the application of the same pulse and recorded the current response, as shown in Fig.\ref{figure3}\,(a). Notably, even after a week, the current response exhibited excellent stability, showing no discernible increase in the probe current. This suggests negligible polarization loss during the testing period and long-term polarization retention.



High fidelity and flexibility of domain engineering are also required for realizing quasi-phase matching in quadradic nonlinear optical devices. We first designed and patterned a "YALE"-shape top electrode, consisting of the four letters which are connected through a bus electrode. The bus is connected to a 20\textmu m$\times$20\textmu m square electrode for convenient contact with the probe tip. Multiple voltage pulses with 62\,V peak voltage and 500\,$\upmu$s length were applied to the contact electrode at room temperature to ensure thorough polarization switching. After poling, the N-polar domains were more susceptible to HCl solution and  can be further exposed via HCl etching for up to 20 minutes. The significant contrast between the original and poled domains can be revealed via scanning electron microscopy(SEM), as shown in Fig.~\ref{figure3}(b). The well-defined boundaries and uniformity of the reversed domains provide strong evidence for high quality poling of ScAlN film via lithographically defined patterns.

Finally, to demonstrate periodic poling of ScAlN, electrode arrays were patterned with varying periods $\Lambda$ of challenging small values, which are 2\textmu m, 1\textmu m, 0.8\textmu m, 0.6\textmu m and 0.4\textmu m, respectively. The same poling voltage pulses used previously were applied to these electrodes, followed by the same post-processing steps. The resulting exposed domains are presented in Fig.~\ref{figure3}(c), where the poling periods agrees well with our designed electrode patterns. The duty cycles for electrode arrays were designed and fabricated to be 25$\%$ for all the periods, but here the achieved poling duty cycles were 31$\%$, 35$\%$, 41$\%$, 54$\%$ and 75$\%$ as period decreases, which we believe can be mediated by optimizing pulse length and electrode width for smaller periods accordingly. Both the smallest poling period and poling uniformity is comparable to that achieved in the state-of-art LiNbO3 platform with lithographically defined electrode \cite{LN_submicron_bipolar}. We hereby demonstrate our capability of realizing periodic poling of thin film ScAlN with arbitrary periods, especially sub-micron periods, which in principle can fullfill quasi-phase matching for $\chi^{(2)}$ frequency conversion in any wavelength range within ScAlN's transparency window. Importantly, the submicron poling period will unlock possibility for mirrorless optical parametric oscillation, which since its first proposal in 1966 \cite{MOPO_1966}, hasn't been verified in any integrated photonic platforms.  %

In conclusion, ScAlN is a promising material for photonic applications due to its flexibility in domian engineering and high fidelity periodical poling reported here. The decent coercive field ~6MV/cm make it possible to pole thicker ScAlN film which provides better optical confinement when patterned into waveguides. Photonic waveguide based on 500nm sputtered ScAlN has been reported before with propagation loss 9±2dB/cm at 1550nm \cite{ScAlN_circuit}, which still holds room for fabrication optimization to achieve comparable loss as AlN photonic circuits \cite{high_fidelity_wet_etch,alex_shg,xianwen_comb}. With future development and growth optimization on waveguide-compatible films, it is feasible to realize low-loss ScAlN photonic circuits for a wide range of integrated $\chi^{(2)}$ nonlinear optics applications, and bring significant advances to III-nitride photonics\cite{liu2023aluminum}. 

\newpage

\vspace{2 mm}
\section*{Acknowledgment} This project is supported in part by Semiconductor Research Corporation (SRC), Defense Advanced Research Projects Agency (DARPA) under the COmpact Front-end Filters at the ElEment-level (COFFEE). FY and HXT acknowledge partial funding from National Science Foundation (NSF) Center for Quantum network (CQN) under grant number EEC-1941583. The facilities used for device fabrication were supported by the Yale SEAS Cleanroom and the Yale Institute for Nanoscience and Quantum Engineering (YINQE). The authors would like to express their gratitude Dr. Yong Sun, Dr. Michael Rooks, Sean Rinehart, and Kelly Woods for their invaluable assistance in device fabrication.

\section*{Data availability} The data that  support the findings of this study are available from the corresponding authors upon reasonable request.

\vspace{2 mm}
\textbf{REFERENCES}
\bibliography{ScAlNpoling}

\end{document}